\documentclass[12pt]{article}
\usepackage{graphicx}
\begin{document}
\begin{center}
{\large\bf Void or Dark Energy?} \vskip 0.3 true in {\large J. W. Moffat} \vskip 0.3 true
in {\it The Perimeter Institute for Theoretical Physics, Waterloo,
Ontario, N2L 2Y5, Canada} \vskip 0.3 true in and \vskip 0.3 true
in {\it Department of Physics, University of Waterloo, Waterloo,
Ontario N2L 3G1, Canada}
\end{center}

\begin{abstract}%
 Two possible explanations for the type SNe Ia supernovae observations are a nonlinear, underdense void embedded in a matter-dominated Einstein-de Sitter spacetime or dark energy in the $\Lambda$CDM model. Both of these alternatives are faced with Copernican fine-tuning problems. A case is made for the void scenario that avoids introducing undetected dark energy.
\end{abstract}
\vskip 0.2 true in e-mail: john.moffat@utoronto.ca


\section{Introduction}

The standard $\Lambda$CDM model is based on Einstein's gravitational theory (GR) and the cosmological principle: there is no special place in the universe - the universe is isotropic and homogeneous. The assumptions of an isotropic and homogeneous universe were used to develop the Lema\^{i}tre-Friedmann-Robertson-Walker (LFRW) expanding universe model, which forms the basis of the standard modern $\Lambda$CDM cosmology. Although this model has successfully described the available cosmological observations, in particular, the WMAP CMB data~\cite{Komatsu}, it has about 8 free adjustable parameters and relies heavily on the simplest assumption of maximal symmetry, the LFRW geometry of spacetime and that GR is the correct theory of gravity to describe the large scale structure of the universe.

A disturbing feature of the standard $\Lambda$CDM model is that roughly 96 percent of the universe is invisible. Approximately 30 percent is composed of ``dark matter", while 70 percent is the dark energy that pervades the universe like a modern-day ether, and is responsible for the asserted acceleration of the expansion of the universe invoked to explain the supernovae observations~\cite{Perlmutter,Riess,Kowalski}. The dark matter has been postulated to explain the rotation curve data of galaxies, the stability of clusters of galaxies, merging clusters such as the Bullet Cluster, gravitational lensing and the WMAP data. After several years of searching for the ubiquitous dark matter, no successful detection of dark matter particles has been achieved. By its nature the dark matter particles interact only with gravity, so the existence of dark matter is inferred through gravitational observations such as galaxy rotation curves.

The dark energy is a mysterious and undetectable uniform substance that has been identified with negative pressure vacuum energy. This interpretation falls prey to the serious lack of understanding of vacuum density in quantum field theory, leading to preposterous degrees of fine-tuning to agree with the ``observed" vacuum density associated with the cosmological constant $\Lambda$. To avoid this problem, many modified gravity theories have been proposed that devote themselves to explaining the accelerated expansion of the universe. Many of these theories suffer from maladies that render them unphysical. They can possess ghosts and instabilities and not be able to explain the precise relativistic corrections engendered by GR in the solar system, such as the Cassini spacecraft observation of the Eddington-Robertson PPN parameter $\gamma -1= (2.1\pm 2.3)\times 10^{-5}$~\cite{Bertotti}.

An early extensive study of cosmological voids~\cite{Moffat,Moffat2} in the exact Lema\^{i}tre-Tolman-Bondi (LTB) inhomogeneous solution of GR~\cite{Lemaitre,Tolman,Bondi,Bonnor,Moffat3,Moffat4,Krasinski}, predicted that the luminosity distance in a void solution would depend on the red shift $z$ in a way that deviates from the LFRW model prediction~\cite{Moffat,Moffat2}. The distance modulus calculated from the void model in 1994-95 suggested an {\it apparent} acceleration of the expansion of the universe as inferred by an observer in an FLRW spacetime~\cite{Moffat2}. Indeed, four years before the celebrated supernovae SNe Ia observations, it was clear that {\it if the inhomogeneous void scenario is correct}, then the observed dimming of the supernovae light was to be expected. However, the void solution demanded that the observer be situated close to the center of the void to maintain the observed isotropy of the large scale structure of the universe. This would imply a violation of the cosmological Copernican Principle. The adherence to a dark energy such as the vacuum energy in the $\Lambda$CDM model, also suffers from an anti-Copernican principle fine-tuning in coordinate time. The negative pressure dark energy and the associated acceleration of the expansion of the universe began dominating the evolution of the standard model when life first appeared on our planet. This is referred to as the ``coincidence" problem. Thus, we are faced with having to make a choice between two theoretical evils, the dark energy scenario or the more conservative inhomogeneous void scenario. As has been recently demonstrated~\cite{Bassett,Clifton,Caldwell,Clarkson,Tomita}, it is difficult with the presently available observational data to distinguish between the idealistic Copernican Principle cosmology and those cosmological models that violate it.

One ongoing vexing problem with the postulate of a cosmological constant in the $\Lambda$CDM standard model is our lack of understanding of the vacuum density in physics. The notorious cosmological constant problem remains unresolved, leading to a huge fine-tuning problem. For example, in the standard electroweak model with a Higgs particle the predicted vacuum density is of order $10^{56}$ times bigger than the vacuum density required in the $\Lambda$CDM model~\cite{Higgs}. The problem becomes even more pronounced when energy scales reach the Planck energy $\sim 10^{19}$ GeV, resulting in a fine-tuning of order $10^{122}$ compared to the ``observed" cosmological constant.

A relativistic modified gravity (MOG) theory known as the Scalar-Tensor-Vector Gravity (STVG) theory has been applied successfully to fit astrophysical as well as cosmological data~\cite{Moffat5,Moffat6,MoffToth,MoffToth2,MoffToth3,MoffToth4,MoffToth5,Brownstein} without non-baryonic dark matter. Thus, if exotic dark matter remains undetected (as was the nineteenth century ether), then to explain the extensive astrophysical and cosmological data, we are forced to modify Newtonian and Einstein gravity. In~\cite{MoffToth2,MoffToth3}, the MOG was able to also explain the accelerated expansion of the universe without a cosmological constant, although this explanation was also subject to the fine-tuned coincidence problem.

In contrast to this need to modify gravity to avoid postulating an undetected dark matter, the void solution to the SNe Ia supernovae observations does not require a modification of Einstein's gravity theory due to its claim that there is no ``dark energy''. An underdense void expands faster than its more dense surrounding galaxies, whereby younger supernovae inside the void would be observed to be receding more rapidly than older supernovae outside the void. By assuming the LFRW homogeneity, the mistaken conclusion would be reached that the expansion rate of the universe is accelerating, while in fact the local void and the global universe are actually decelerating. The local void scenario has been investigated by many authors. For a review, see~\cite{Celerier} and for more recent citations see~\cite{Sarkar}. There are other problems occurring in the void model, in addition to giving up the cosmological Copernican Principle. To fit the CMB data and other pertinent cosmological data, the radial size of the void required to fit the supernovae data is hundreds of Mpc to Gpc, and we must be close to the center of the void to avoid an unacceptably large CMB dipole~\cite{Moffat,Alnes}.

Another problem to be considered in the inhomogeneous void model is the assumed initial conditions of the universe. Strictly speaking, simple models of inflation would be inconsistent with the existence of a large inhomogeneous void. However, there exists an alternative solution to the initial value horizon and flatness problems. This is based on a bimetric gravity variable speed of light (VSL) model~\cite{Moffat3}. It has been demonstrated by Magueijo~\cite{Magueijo} that this model in conjunction with the Dirac-Born-Infeld model can predict almost scale invariant primordial fluctuations in agreement with observations without inflation. This model would not necessarily negate the possibility of large, nonlinear and inhomogeneous voids.

In the following, we will review discoveries made in the early investigations of the void model~\cite{Moffat,Moffat2}, and consider the age problem in cosmology and how the inhomogeneous cosmology can ameliorate this problem. We also consider the possible amplification of growth structure and its early onset in the inhomogeneous LTB model. We do not consider the important issue of confronting the void model with the Cosmic Microwave Background (CMB) WMAP data~\cite{Komatsu}. Attempts to do this have been made in the recent literature~\cite{Sarkar,Tomita2,Tomita3,Biswas,Zibin,Garcia}. These investigations remain inconclusive as to how well the void model can produce the acoustical power spectrum, the matter power spectrum associated with large scale surveys of galaxies and the observed baryon acoustical oscillations (BAO) in the correlation function. The role of structure growth in the inhomogeneous models plays an important role in these cosmological calculations, as does the importance of contributions from ``warm" dark matter such as neutrinos with a mass not violating recent upper bounds obtained from WMAP observations. We conclude that the void model has not been ruled out. However, the best fits to the CMB WMAP data and the BAO appear to require a low value of $H_0\sim 44\,{\rm km\,s^{-1}\,Mpc^{-1}}$ compared to measurements of $H_0$~\cite{Jackson}. We should nonetheless be cautious about the true value of $H_0$, for it has a long history of changing its value with time. Moreover, in an inhomogeneous cosmology the value of the Hubble constant is position dependent. The local measured value of $H_0$ near the center of the void is expected to be larger than in distant parts of the universe outside the void. This could accommodate fits to the CMB data at the surface of last scattering with lower values of $H_0$ than would be measured by an observer close to the center of the void.

\section{Inhomogeneous Friedmann Equations}

For the sake of notational clarity, we write the FLRW line element
\begin{equation}
ds^2=dt^2-a^2(t)\biggl(\frac{dr^2}{1-kr^2}+r^2d\Omega^2\biggr),
\end{equation}
where $k=+1,0,-1$ for a closed, flat and open universe,
respectively, and $d\Omega^2=d\theta^2+\sin^2\theta d\phi^2$. The
general, spherically symmetric inhomogeneous line element is given
by~\cite{Lemaitre,Tolman,Bondi,Bonnor,Moffat,Moffat2,Moffat3,Moffat4,Krasinski}:
\begin{equation}
\label{inhomometric} ds^2=dt^2-X^2(r,t)dr^2-R^2(r,t)d\Omega^2.
\end{equation}
The energy-momentum tensor ${T^\mu}_\nu$ takes the barytropic form
\begin{equation}
\label{energymomentum} {T^\mu}_\nu=(\rho+p)u^\mu u_\nu
-p{\delta^\mu}_\nu,
\end{equation}
where $u^\mu=dx^\mu/ds$ and, in general, the density
$\rho=\rho(r,t)$ and the pressure $p=p(r,t)$ depend on both $r$
and $t$. We have for comoving coordinates $u^0=1, u^i=0,\,
(i=1,2,3)$ and $g^{\mu\nu}u_\mu u_\nu=1$.

The Einstein gravitational equations are
\begin{equation}
\label{Einstein} G_{\mu\nu}+\Lambda g_{\mu\nu}=-8\pi GT_{\mu\nu},
\end{equation}
where $G_{\mu\nu}=R_{\mu\nu}-\frac{1}{2}g_{\mu\nu}{\cal R}$,
${\cal R}=g^{\mu\nu}R_{\mu\nu}$ and $\Lambda$ is the cosmological
constant. Solving the $G_{01}=0$ equation for the metric
(\ref{inhomometric}), we find that
\begin{equation}
X(r,t)=\frac{R'(r,t)}{f(r)},
\end{equation}
where $R'=\partial R/\partial r$ and $f(r)$ is an arbitrary
function of $r$.

We obtain the two generalized Friedmann equations~\cite{Moffat}:
\begin{equation}
\label{inhomoFriedmann} \frac{{\dot R}^2}{R^2}+2\frac{{\dot
R}'}{R'}\frac{{\dot R}}{R}+\frac{1}{R^2}(1-f^2)
-2\frac{ff'}{R'R}=8\pi G\rho+\Lambda,
\end{equation}
\begin{equation}
\label{inhomoFriedmann2} \frac{\ddot
R}{R}+\frac{1}{3}\frac{\dot{R}^2}{R^2}
+\frac{1}{3}\frac{1}{R^2}(1-f^2) -\frac{1}{3}\frac{{\dot
R}'}{R'}\frac{{\dot R}}{R}+\frac{1}{3} \frac{ff'}{R'R}
=-\frac{4\pi G}{3}(\rho+3p)+\frac{1}{3}\Lambda,
\end{equation}
where $\dot R=\partial R/\partial t$.

Consider now the Lema\^{\i}tre-Tolman--Bondi model~\cite{Lemaitre,Tolman,Bondi} for
a spherically symmetric inhomogeneous universe filled with dust. The line
element in comoving coordinates can be written as:
\begin{equation} \label{linel}
 ds^{2}=dt^{2}-R^{\prime 2}(t,r)f^{-2}dr^{2}-R^{2}(t,r)d\Omega^{2},
\end{equation}
where $f$ is an arbitrary function of $r$ only, and the field equations
demand that $R(t,r)$ satisfies:
\begin{equation} \label{F}
 2R\dot{R}^{2}+2R(1-f^{2})=F(r),
\end{equation}
with $F$ being an arbitrary function of class $C^{2}$. We have three distinct
solutions depending on whether $f^{2}<1$, $=1$, $>1$ and they
correspond to elliptic (closed), parabolic (flat) and hyperbolic (open) cases,
respectively. The proper density can be expressed as:
\begin{equation} \label{dens}
 \rho=\frac{F^{\prime}}{16\pi R^{\prime} R^{2}}.
\end{equation}

The total mass within comoving radius $r$
is given by
\begin{equation} \label{mass}
M(r)= \frac{1}{4} \int_{0}^{r} dr f^{-1} F^{\prime} =4 \pi
\int_{0}^{r} dr \rho f^{-1} R^{\prime} R^{2},
\end{equation}
so that
\begin{equation}
 M^{\prime} (r) = \frac{dM}{dr} = 4 \pi \rho f^{-1} R^{\prime}
R^{2}.
\end{equation}
Also for $\rho > 0$ everywhere we have $F^{\prime} >0$ and
$R^{\prime} >0$ so that in the non-singular part of the model $R>0$
except for $r=0$ and $F(r)$ is non-negative and monotonically
increasing for $r \geq 0$. This could be used to define the new
radial coordinate $\bar{r}^{3}=M(r)$ and find the parametric
solutions for the rate of expansion.

For the closed and open cases the parametric solutions for the rate
of expansion can be written as
\begin{equation}
\label{parneg}
R=\frac{1}{4} F {\left( 1 - f^{2} \right)}^{-1} \left[1-\cos(v)
\right] ,\quad    f^{2}<1 ,
\end{equation}
\begin{equation}
 t+\beta= \frac{1}{4} F {\left( 1-f^{2} \right)}^{-3/2}\left[ v-
\sin(v) \right]   ,\quad f^{2}<1 ,
\end{equation}
and
\begin{equation}
\label{parpos}
R= \frac{1}{4} F {\left( f^{2}-1\right)}^{-1} \left[\cosh(v)-1
\right] ,\quad   f^{2}>1 ,
\end{equation}
\begin{equation}
\label{parpost}
 t+\beta= \frac{1}{4} F {\left( f^{2}-1\right)}^{-3/2} \left[
\sinh(v)-v \right]   ,\quad f^{2}>1 ,
\end{equation}
with $\beta(r)$ being again a function of integration of class
$C^{2}$ and $v$ the parameter.

In the flat (parabolic) case $f^{2}=1$, we have
\begin{equation}
R = \frac{1}{2}{\left(9F\right)}^{1/3}{\left(t+\beta\right)}^{2/3},
\end{equation}
with $\beta(r)$ being an arbitrary function of class $C^{2}$ for
all $r$. After the change of coordinates $R(t,\bar{r}) = \bar{r}
{\left(t+\beta(\bar{r})\right)}^{2/3}$, the metric becomes:
\begin{equation} \label{metric}
 ds^{2}=dt^{2}-{\left(t+ \beta \right)}^{4/3} \left( Y^{2}
  dr^{2}+r^{2}d\Omega^{2} \right),
\end{equation}
where
\begin{equation}
 Y= 1 + \frac{2 r {\beta}^{\prime}}{3 \left(t+ \beta \right)},
\end{equation}
and from (\ref{dens}) the density is given by
\begin{equation} \label{densb}
 \rho = \frac{1}{6 \pi {\left(t+ \beta \right)}^{2} Y} .
\end{equation}
Clearly, we have for finite $\beta^{\prime}$ that for $t \rightarrow \infty$ the model tends to
the flat Einstein--de Sitter case.

The flat case ($f^{2}=1$) has been extensively studied~\cite{Moffat}.
The model depends on one arbitrary function $\beta (r)$ and could be
specified by assuming the density on some space-like hypersurface, say
$t=t_{0}$. However, specifying the density on the past light cone of the
observer is more appropriate.

Before we proceed to discuss the observational grounds
for modeling a local void, we need to amplify the discussion of the LTB
model by introducing basic features of the propagation of light. The high
degree of isotropy of the microwave background forces us to the conclusion
that we must be located close to the spatial center of the local LTB void. We place an observer at the center
($t_{\rm obs}=t_{0} , r_{\rm obs}=0$).

The luminosity distance between an observer at the origin of our
coordinate system ($t_{0},0$) and the source at
($t_{e},r_{e},\theta_{e},\phi_{e}$) is given by
\begin{equation} \label{lumdis}
 d_{L}={\left(\frac{{\cal L}}{4\pi{\cal F}}\right)}^{1/2}=R(t_{e},r_{e}){\left[1
 + z(t_{e},r_{e}) \right]}^2,
\end{equation}
where ${\cal L}$ is the absolute luminosity of the source (the
energy emitted per unit time in the source's rest frame), ${\cal
F}$ is the measured flux (the energy per unit time per unit area as
measured by the observer) and $z(t_{e},r_{e})$ is the redshift
(blueshift) for a light ray emitted at ($t_{e},r_{e}$) and observed
at ($t_{0},0$).

The light ray traveling inwards to the center satisfies:
\begin{equation}
 ds^{2}=dt^{2}-R^{\prime 2}(t,r)f^{-2}dr^{2}=0 ,\quad
d\theta=d\phi=0,
\end{equation}
and thus
\begin{equation}
\label{zero}
\frac{dt}{dr}=-R^{\prime}(t,r)/f(r),
\end{equation}
where the sign is determined by the fact that the light ray travels along the
{\em past} light cone (i.e. if $r_{e'}>r_{e''}$, then $t_{e'}<t_{e''}$).

If the equation of the light ray traveling along the light cone is
\begin{equation} \label{rays}
 t = T(r) ,
\end{equation}
we get the equation of a ray along its path:
\begin{equation} \label{lightcone}
 \frac{dT(r)}{dr}=-\frac{R^{\prime}}{f} [T(r),r].
\end{equation}
Consider two rays emitted by the source with the small time separation $\tau$.
The equation of the first ray is
\begin{equation} \label{morerays}
 t = T(r) ,
\end{equation}
while the equation of the second ray is
\begin{equation}
t=T(r)+\tau(r).
\end{equation}
Using Eq.(\ref{zero}) we get the equation of a ray and the rate of change of $\tau(r)$ along the path:
\begin{equation}
\frac{dT(r)}{dr}=-R^{\prime}[T(r),r],
\end{equation}
and
\begin{equation}
\frac{d\tau(r)}{dr}=-\tau(r)\dot{R}^{\prime}[T(r),r],
\end{equation}
where
\begin{equation}
{\dot{R}}^{\prime}[T(r),r] =
 {\left. \frac{{\partial}^{2}R}{\partial t \partial r} \right
 |}_{r,T(r)}={\left. \frac{\partial R^{\prime}}{\partial t}
 \right |}_{r,T(r)} .
\end{equation}
If we take $\tau(r_e)$ to be the period of some spectral line at $r_e$ then
\begin{equation}
\frac{\tau(0)}{\tau(r_e)}=\frac{\nu(r_e)}{\nu(0)}=1+z(r_e),\quad z=0\quad {\rm for} \quad r_e=0.
\end{equation}
The equation for the redshift considered as a function of $r$ along
the light cone is:
\begin{equation} \label{zred}
 \frac{dz}{dr}=(1+z){\dot{R}}^{\prime}[T(r),r].
\end{equation}
The shift $z_1$ for a light ray traveling from $(t_1,r_1)$ to $(t_0,0)$ is given by
\begin{equation}
\label{lnequation}
\ln(1+z_1)=-\int^{r_1}_0 dr \dot{R}^{\prime}[T(r),r].
\end{equation}

The shift $z_{1}$ for a light ray traveling from ($t_{1},r_{1}$) to
($t_{0},0$) is to linear order:
\begin{equation} \label{grco}
  \ln (1+z_{1})  =  - \ln (1-a_{1}) - \int_{0}^{r_{1}} dr
\frac{M^{\prime}(r)}{r(1-a_{1})} ,
\end{equation}
where
\begin{equation}
a_{1}(r)=\dot{R}[T(r),r].
\end{equation}
In obtaining equation (\ref{grco}), we used (\ref{dens}) and
(\ref{mass}). Thus we have two contributions to the redshift: the
cosmological redshift due to expansion, described by the first term with
$a_{1}=\dot{R}$, and the gravitational shift due to the difference between
the potential energy per unit mass at the source and at the observer.
In the homogeneous case $M^{\prime}(r)=0$, so there is no
gravitational shift.

\section{Local Void} \label{void}

If we restrict ourselves to spatial scales that have been well probed
observationally, i.e. up to a few hundred Mpc, the most striking feature of
the luminous matter distribution is the existence of large voids surrounded
by sheet-like structures containing galaxies. Early surveys
give a typical size of the voids of the order 50--60 $h^{-1}$ Mpc~\cite{Geller,Hoyle}.

We study a void with the central density equal to that of an LFRW model
with the density parameter $\Omega_{0}=0.2$, asymptotically approaching
the Einstein-de Sitter model with $\Omega_{0}=1$~\cite{Moffat,Moffat2}.
Since cosmological observations are done by detecting some form of electromagnetic
radiation, the mapping obtained from them describes the density along the light cone.
We describe the density distribution, as a function of the redshift $z$, by:
\begin{equation}
\label{densdistr}
\Omega_{\rm void}(z)=\frac{\Omega_{min}+{(z/a)}^2}{\Omega_{max}+{(z/a)}^2}.
\end{equation}
Choosing a rational function $\Omega_{\rm void(z)}$ as opposed to, say, a
gaussian distribution does not affect significantly the resulting calculation. In the numerical calculations~\cite{Moffat2}, we used
the values $\Omega_{min}=0.2$ and $\Omega_{max}=1$. The two density distributions are parameterized by (A) $a=0.125$ and (B)
$a=0.25$, and are depicted in Figure 1.

\begin{figure}[ht]
\includegraphics[height=\linewidth,angle=270]{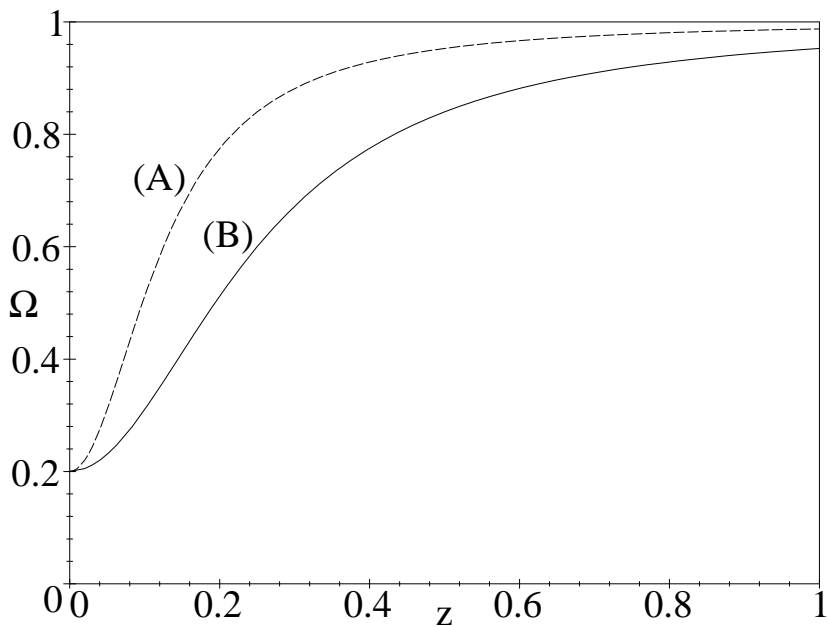}
\caption{The density distributions $\Omega_{\rm void}$ as functions of red shift. The cases (A) and (B) are described in the text.}
\end{figure}

As seen from (\ref{linel}) and (\ref{dens}) shell crossing can occur when
$R^{\prime}(t,r)=0$ for some $r=r_s$. This, in general, may lead to
$\lim_{r \rightarrow r_s} \rho = \infty$ and a shell crossing singularity.
Different shells $r=\mbox{const.}$ collide and the comoving coordinates
become inadmissible. The singularity can be avoided, if the functions
$F^\prime(r)$ and $f(r)$ both have at $r=r_s$ zeros of the same order as
$R^\prime(t,r)$ has. This, however, is not the case in the present model, for we
assume $F(r) \geq 0$ and $F^\prime(r) \geq 0$.

In the model shell crossing does occur. The physical
significance of this can be viewed from two different points of view.
One is to regard a shell crossing singularity as unphysical and try
to exclude it by assuming special initial conditions. However, this leads to
difficulties with adjusting the model to observational data. Either
the average density of the void is not appreciably lower than the density of
the background, or the matter in the void is much older than outside. We will treat the shell crossing
singularity as a physical phenomenon, viewing the shell crossings as surface layers of matter, interpreted as walls of the
expanding void corresponding to the sheet-like structures surrounding observed voids.

For the distribution (\ref{densdistr}), the shell crossing happens in
the distant future. The LTB void studied
here is applicable to the matter dominated era. The fact that the LTB model
does not allow for pressure is not a serious problem. The universe has been matter
dominated since $z \approx 10^4$. The
model is applicable to this era and may be thought of as a continuation of
an earlier --very nearly-- LFRW model, provided its density contrast at the
time of decoupling, $t=t_{dc}$, is $\rho(t_{dc},r) / \rho_{LFRW}(t_{dc})
\approx1$.)

Figure 2 depicts the logarithmic redshift--luminosity distance relations
for both LTB voids and both time scales as published in 1994-95~\cite{Moffat2}. A survey of galaxy
measurements relevant in the early nineties is also included~\cite{Lilley}. The $z(d_L)$ relations for the LFRW cases with $\Omega_0=1$ and $\Omega_0 = 0.2$ are presented for comparison.

\begin{figure}[ht]
\begin{minipage}{0.47\linewidth}\includegraphics[height=1.1\linewidth,angle=270]{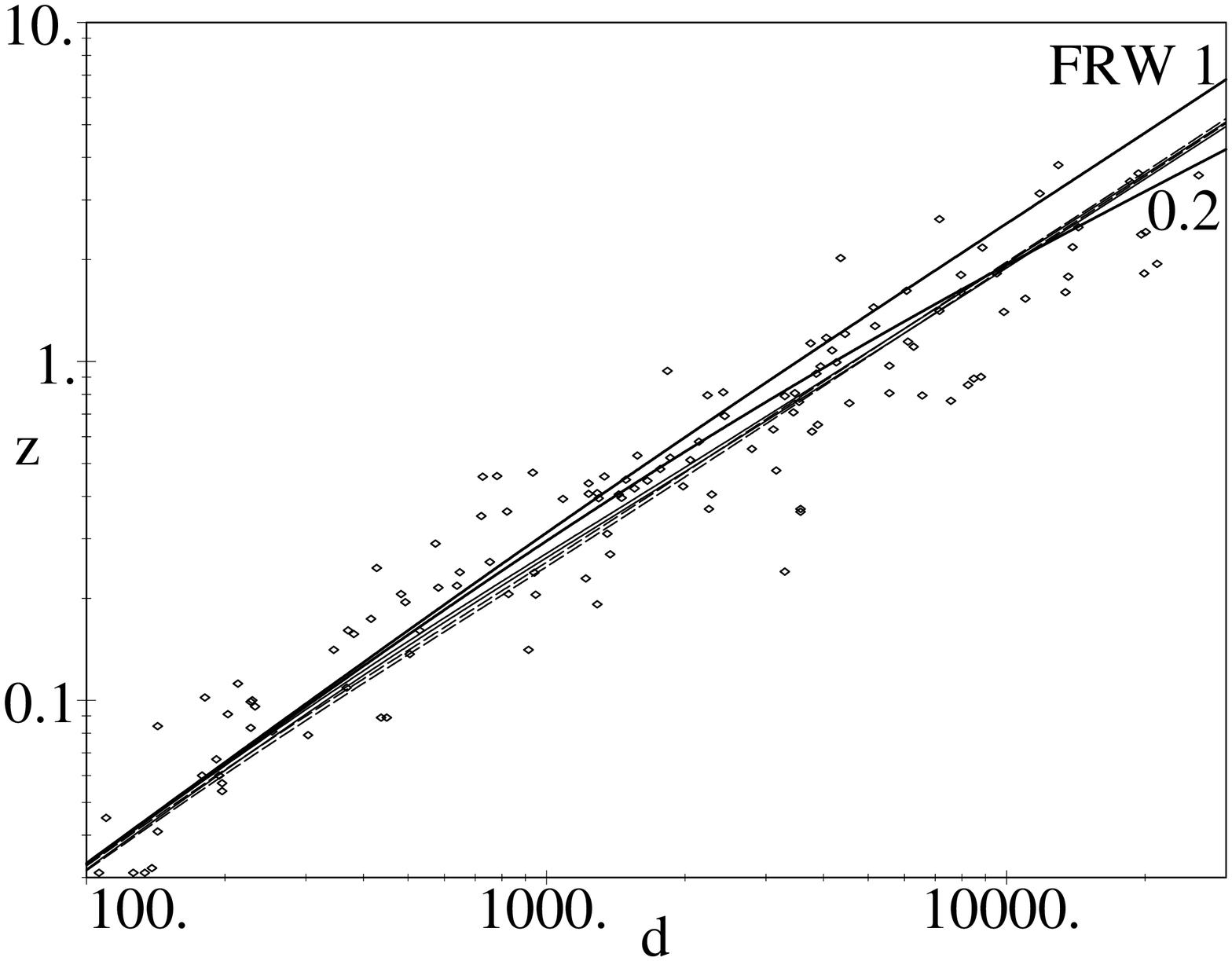}
\end{minipage}~\begin{minipage}{0.47\linewidth}~\vskip 0.3in\includegraphics[width=0.98\linewidth]{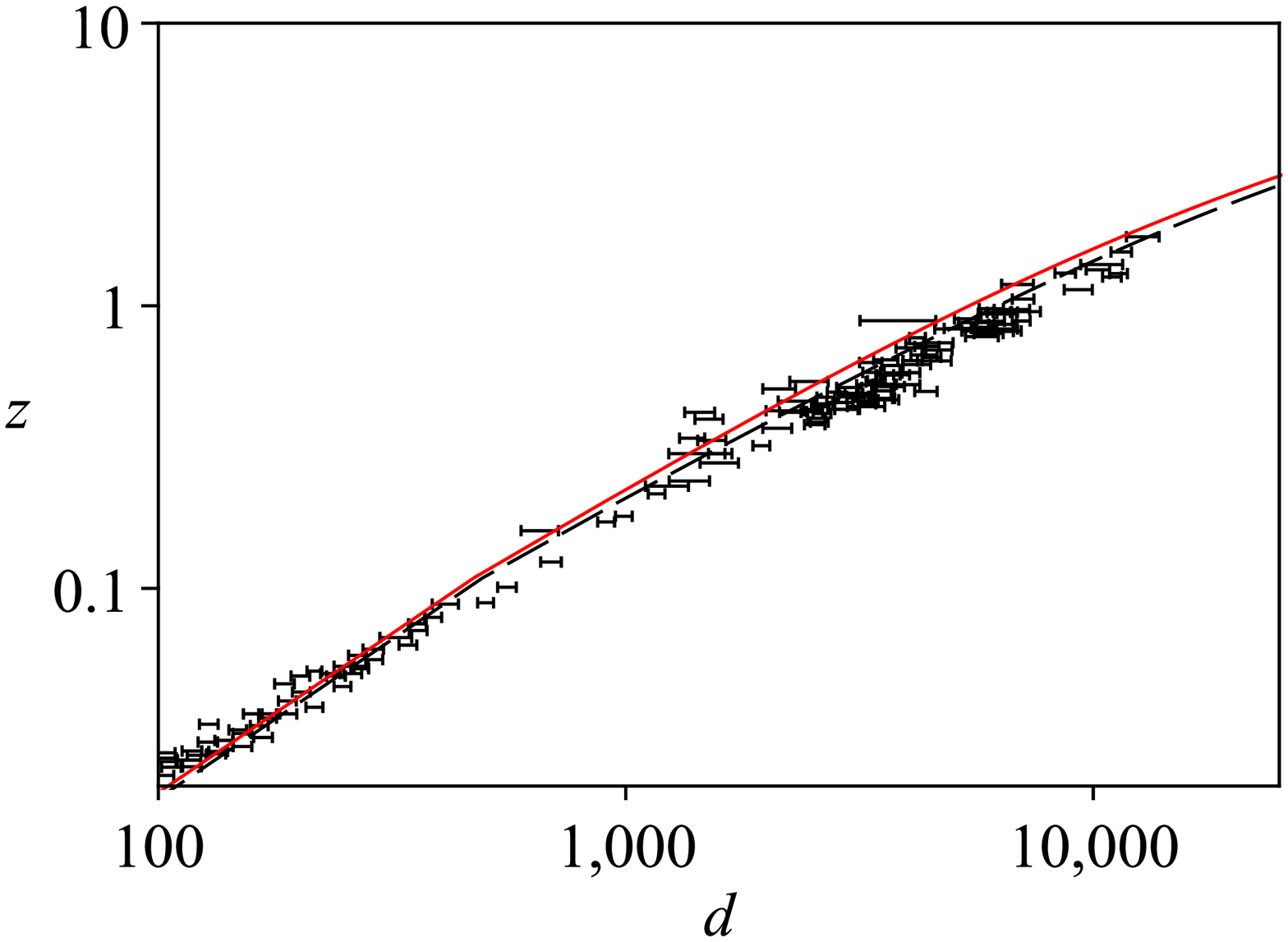}
\end{minipage}
\caption{The $\log(z)$ vs. $\log(d_L)$ relations.
In the left-hand figure~\cite{Moffat2}, observational data, denoted by $\diamond$, are adapted from~\cite{Lilley} and
the luminosity distance $d_L$ is given in Mpc. In the right-hand figure, the LFRW result for $\Omega_0=1$ is shown as a solid red curve, while the $\Lambda$CDM result with $\Omega_M=0.27$ and $\Omega_\Lambda=0.73$ is shown as a dashed black curve. SN 1a data are shown for comparison.}
\end{figure}

In the standard LFRW cosmology, we have in the matter-dominated era:
\begin{equation}
\Omega_\Lambda+\Omega_M+\Omega_K=1,
\end{equation}
where
\begin{equation}
\Omega_\Lambda=\frac{8\pi G\rho_V}{3H_0^2},\quad \Omega_M=\frac{8\pi G\rho_M}{3H_0^2},\quad
\Omega_K=\frac{K}{a_0^2H_0^2},
\end{equation}
where $\rho_V, \rho_M$ and $K$ denote the vacuum density, the matter density and the curvature constant $K$, respectively. For the flat space standard $\Lambda$DCM cosmology we have $\Omega_K=0$ and the luminosity distance of a source with redshift $z$ is
\begin{equation}
d_L=a_0r_1(1+z)=\frac{1+z}{H_0}\int^1_{\frac{1}{1+z}}\frac{dx}{\sqrt{x^2(\Omega_\Lambda+\Omega_Mx^{-3})}}.
\end{equation}
The distance modulus is the difference between the apparent magnitude $m$ and the absolute magnitude $M$, given by
\begin{equation}
\label{distmodulus}
\mu\equiv m-M=5\log_{10}\biggl(\frac{d_L}{1\,{\rm Mpc}}\biggr)+25,
\end{equation}
where
\begin{equation}
d_L=(1+z)^2d_A,
\end{equation}
and $d_A$ is the angular diameter distance.

We see from Fig.2 and Eq.(\ref{distmodulus}) that already four years before the type SNe Ia supernovae measurements were published~\cite{Perlmutter,Riess}, it was clear that if the void model was correct, {\it then the predicted deviations from the LFRW model would lead to unexpected results for red shifts $0.2 < z < 2$, namely, an apparent dimming of the supernovae light, and an apparent acceleration of the expansion of the universe as inferred by an observer using the LFRW model.} Several recent papers have shown that if the void has a sufficiently large radius, then excellent fits to the supernovae data can be obtained from the LTB void model~\cite{Alnes,Biswas,Iguchi,Mansouri,Yoo}.

Let us now consider the observational properties of our
model with a local LTB void by concentrating on the Hubble parameter
measurements. In a spherically symmetric model, we have two
``Hubble parameters'': $H_{r}(t,r)$ for the local expansion rate in the radial
direction and $H_{\bot}(t,r)$ for expansion in the perpendicular direction.
Usual definitions give~\cite{Moffat,Moffat2}:
\begin{equation}
\label{hubble}
H_{r}=\frac{\dot{l}_{r}}{l_{r}}=\frac{\dot{R}^{\prime}}{R^{\prime
}},
\end{equation}
\begin{equation} \label{hubblep}
H_{\bot}=\frac{\dot{l}_{\bot}}{l_{\bot}}=\frac{\dot{R}}{R},
\end{equation}
where $l$ denotes the proper distance, i.e. $l_{r} = R^{\prime} (t,r) f^{-1}
dr$ and $l_{\bot} = R(t,r) d\Omega$. Due to the fact that there are both
gravitational and expansion redshifts contributing to the total $z$, neither of
the Hubble parameters $H_r$, $H_{\bot}$ is fully analogous to the LFRW's
$H_{LFRW}=\dot{a}/a$. The closest analogy exists for small separations of
the source and the observer. Using (\ref{hubblep}) and (\ref{lumdis}) we get for small
$r_{e}$:
\begin{equation}
 z (t_{e},r_{e})=H_{\bot}(t_{e},r_{e}) d_{L} (t_{e},r_{e}),
\end{equation}
which is formally analogous to the LFRW result. Two main differences are
that our relation is {\em local} and that from cosmological observations (on
small scales) we obtain the angular Hubble parameter $H_{\bot}=\dot{R} /
R$ rather than $H_{LFRW}=\dot{a}/a$.

In general, if we lived in a local LTB void and the $z$ versus $d_L$ relation
differed from the LFRW one, but we were biased
by our theoretical prejudice and interpreted cosmological observations
through an LFRW model, we would expect the value of the Hubble
parameter to be position and $d_L$ dependent.

\begin{figure}[ht]
\includegraphics[height=\linewidth,angle=270]{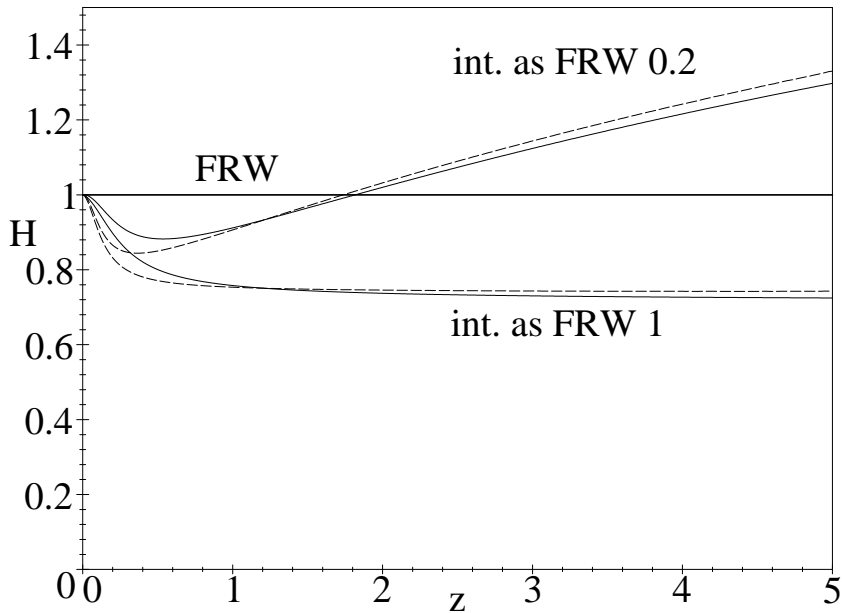}
\caption{The ``observed" Hubble constant $H$ in units of the local measurement of $H_0$ as a function
of the redshift $z$ from ref.~\cite{Moffat2}.}
\end{figure}

Let us recall that in LFRW cosmology the exact
result for the Hubble relation $d_L$ versus $z$ in the matter dominated
universe is
\begin{equation}
\label{hubblefrw}
d_L =\frac{1}{H_0q_0^2} \left[ zq_0 + \left(q_0-1\right)\left(\sqrt{2zq_0+1} -1
\right) \right],
\end{equation}
where $q_0 \equiv -\ddot{a}(t_0)/a(t_0){H_0}^2$ is the deceleration
parameter.

On cosmologically very small distances, we measure the same value of
$H_0$ independently of the model (we call this value ``the local
measurement''). This stems from the fact that, due to our assumptions, very
close to the center ($r \ll 1$) the model is well approximated by the LFRW
universe with $\Omega=0.2$. Obviously, if the universe were locally LTB
rather than LFRW, then the Hubble parameter based on the observed LTB
values of $z$ and $d_L$, but inferred through an LFRW relation
(\ref{hubblefrw}), would be position (redshift) dependent. The dependence
of the Hubble parameter $H$ (in units of the $H_0$ value as measured
locally) on the redshift $z$ is shown in Figure 3.

The $H(z)$ relation depicted in Figure 3 strongly depends on the
choice of the LFRW case, namely, the choice of the deceleration parameter $q_0$ in
(\ref{hubblefrw}) through which we interpret the observations. The main
feature in the LFRW $\Omega=1$ interpretation is a distinct monotonic
correlation between $H$ and $z$ or $d_L$.  The values of the
``observed'' Hubble constant decrease with the redshift $z$ asymptotically
approaching the background limit. If we interpret the results within the
LFRW $\Omega<1$ framework, the ``observed'' values of the Hubble
constant first decrease with $z$ and then asymptotically increase to some
background limit. The position of the minimum in $H$ depends on the
size of the LTB void.

\section{The Age Problem and Structure Formation}

In the standard $\Lambda$CDM model the assumption that the acceleration of the expansion of the universe is caused by a vacuum dark energy, leads to an increase in the age of the universe. For the flat LFRW model without a cosmological constant $\Lambda$, the present age of the universe at $z=0$ is~\cite{Weinberg}:
\begin{equation}
t_0=\frac{2}{3H_0}=9.3\biggl(\frac{70\,{\rm km/sec/Mpc}}{H_0}\biggr)\,\,{\rm Gyr},
\end{equation}
which is younger than the oldest observed stars and globular clusters in our galaxy. When the vacuum energy is included in the calculation of the age of the universe, we get
\begin{equation}
t_0=(13.4\pm 1.3)\biggl(\frac{70\,{\rm km/sec/Mpc}}{H_0}\biggr)\,\,{\rm Gyr}.
\end{equation}
This is in better agreement with the ages of clusters and stars. In particular, for globular clusters their ages are variously between $11.5\pm 1.3$ Gyr and $14.0\pm 1.2$ Gyr~\cite{Chaboyer,Carretta}. Schramm~\cite{Schramm} gave as the ages of globular clusters $14\pm 2({\rm statistical})\pm 2({\rm systematic})$ Gyr. Even so, the age estimate in the $\Lambda$CDM model {\it is uncomfortably close to the ages of the oldest globular clusters and possibly to the ages of the most distant galaxies observed.}

In our inhomogeneous model, the metric and density are singular on two hypersurfaces:
\begin{equation}
t+\beta=0,\quad Y=0,
\end{equation}
namely,
\begin{equation}
t_1=-\beta,\quad t_2=-\beta-\frac{2r\beta'}{3}.
\end{equation}
The model is valid only for
\begin{equation}
t > \Sigma(r)\equiv {\rm Max}[t_1(r),t_2(r)].
\end{equation}
Here, $t(r)=\Sigma(r)$ defines the big-bang hypersurface in the model. Physically, because the model is pressureless, we interpret $\Sigma(r)$ as the surface on which the universe enters the matter-dominated era. In the LFRW model this occurs at the same time $t_{eq}$ when radiation and matter are equal $t_{eq} \sim 10^4$. However, even in a globally flat inhomogeneous model this can occur at different times. We also note that in the limit $t\rightarrow\infty$, the LTB model gives the Einstein-de Sitter universe:
\begin{equation}
ds^2=dt^2-t^{4/3}(dr^2+r^2d\Omega^2).
\end{equation}

At $r=0$ the big-bang model hypersurface is located at $\Sigma(0)=-\beta(0)$. The requirement that $\beta'(r)$ tends to a finite limit as $r\rightarrow\infty$ forces $\beta'(0)=0$. For an observer at $t(0,0)$ in our void, where $t_0$ is the time coordinate of constant time hypersurface ``now", the age of the universe is given by
\begin{equation}
t_{LTB}=t_0+\beta(0)=\frac{2}{3H_\perp(t_0,0)}.
\end{equation}
Depending on the choice of $\beta(0)$ we can increase the age of the universe as observed by an earth-based observer at $r=0$. If we set $\beta(0)=0$, then
\begin{equation}
t_{LTB}=\frac{2}{3H_\perp(t_0)}=\frac{2}{3H_0}.
\end{equation}
Here, we have replaced the LFRW local value of the Hubble constant $H_0$ by $H_\perp(t_0,0)$. However, for the outer parts of the universe for large $z$ we can choose $\beta(r)$ and $\Sigma(r)$, {\it so that we are able to obtain an age of the universe much more compatible with the ages of globular clusters and radioactive dating and the ages of the most distant galaxies}.

Let us now turn our attention to structure formation in LTB models. Several authors have investigated the growth of structure in the inhomogeneous LTB model~(see, for example, \cite{Tomita4,Clarkson2}). We assume for simplicity that the universe is spatially flat. This is not compatible with our local void scenario but we can extend the present results to a realistic void model. The density contrast in a spatially flat model is described by
\begin{equation}
\delta({\bf x})=\frac{\delta\rho({\bf x})}{\bar\rho}=\frac{\rho(\bf x)-\bar\rho}{\bar\rho},
\end{equation}
where $\bar\rho$ is the average density of the universe. The Fourier transform is
\begin{equation}
\delta_k=V^{-1}\int d^3x\delta(\bf k)\exp(i{\bf k}\cdot{\bf x}).
\end{equation}
The autocorrelation function $\xi({\bf r})$ is defined by
\begin{equation}
\xi({\bf r})=\langle\delta({\bf x}+{\bf r})\delta({\bf x})\rangle.
\end{equation}
We also have
\begin{equation}
\xi({\bf r})=\frac{1}{(2\pi)^2V}\int d^3k\vert\delta_k\vert^2\exp(-i{\bf k}\cdot{\bf x}),\quad \vert\delta_k\vert^2=V\int d^3r\xi({\bf r})\exp(i{\bf k}\cdot{\bf r}).
\end{equation}

In our inhomogeneous and isotropic case (we as an observer are close to the center of the void) $\xi({\bf r})=\xi(r)$ and $\xi(0)=(\delta\rho/\rho)^2$. The density perturbations for the growing mode in the LFRW model obey the equation:
\begin{equation}
\delta_{LFRW}=\delta_{LFRW}(t_{eq})\biggl(\frac{t_{LFRW}}{t_{eq}}\biggr)^{2/3},
\end{equation}
where $t_{eq}$ is the time of radiation and matter equality, $t_{LFRW}$ denotes the time from the initial singularity to a given value of time coordinate $t$, and $t_{LFRW}$ is the same everywhere in the LFRW model.

In our inhomogeneous isotropic flat model we have~\cite{Moffat}:
\begin{equation}
\delta_{LTB}(t,r)=\delta_{LTB}(t_\Sigma,r)\biggl(\frac{t_{LTB}}{t_\Sigma(r)}\biggr)^{2/3},
\end{equation}
where $t_{LTB}(r)=t-\Sigma(r)$ is the time from the initial singularity. We also assume that
\begin{equation}
\delta_{LFRW}(t_{eq})=\delta_{LTB}(t_\Sigma(r),r),
\end{equation}
and $t_{eq}=t_\Sigma(r)$ for all $r$. We now find that there is an amplification of the FLRW perturbation growth in our inhomogeneous model:
\begin{equation}
\delta_{LTB}(t,r)=\biggl(\frac{t_{LTB}(r)}{t_{LFRW}}\biggr)^{2/3}\delta_{LFRW}(t).
\end{equation}
The larger $t_{LTB}(r)$ for a given $r$, the more the structure growth has developed.

For the correlation function we have
\begin{equation}
\xi_{LTB}(t_0,r)=\langle\delta_{LTB}(t_0,r)\delta_{LTB}(t_0,0)\rangle.
\end{equation}
We now have the amplified correlation function in our LTB inhomogeneous model~\cite{Moffat}:
\begin{equation}
\xi_{LTB}=\biggl(\frac{t_{LTB}}{t_{LFRW}}\biggr)^{2/3}\xi_{LFRW}(r).
\end{equation}

The amplification of the structure growth in our model can influence the estimated late-time integrated Sachs-Wolfe (ISW) effect as compared to the standard LFRW model. Moreover, the size of $t_{LTB}$ can increase the time when structure growth of primordial perturbations enter the horizon.

\section{Conclusions}

We have investigated whether an underdense, nearly spherically symmetric void can successfully describe cosmological observations. In spite of the impressive success of the standard $\Lambda$CDM model, based on a homogeneous, isotropic and spatially flat LFRW universe, in fitting a large amount of data including both the type Ia supernovae data and the CMB data, there are problems with the standard model: the undetected dark matter, the cosmological constant problem and the anti-Copernican coincidence problem associated with the vacuum density dark energy. To avoid the generic problems with dark energy the possibility of having a cosmology with a large underdense void, embedded in an Einstein-de Sitter spacetime has been proposed. This avoids having to modify Einstein's gravity theory at least as far as the issue of dark energy is concerned. We have shown that already in an early publication~\cite{Moffat2} the solution with a suitable void density profile, anticipated the supernovae observation of a dimming of the supernovae light, because the void expands faster than the surrounding overdense shell of matter. However, in the void model the acceleration of the universe, as interpreted in an LFRW universe, is a misinterpretation of the observations, for the void and the universe outside the void are in fact decelerating. This means that in the void model we can discard the idea of an undetected dark energy.

The void solution also has the ``philosophical" complication of giving up the cosmological Copernican Principle by requiring that the observer be near the center of the void. Whether cosmology demands a fully Copernican Principle is still an observationally unresolved question. However, the void scenario can only be convincing, if it is fully consistent with the accurate CMB WMAP data, and the observed matter power spectrum based on large scale galaxy surveys, the ISW effect and the BAO data. We cannot rule out the possibility yet that the inhomogeneous void cosmology can fit this data satisfactorily. The question as to how perturbation growth of structure both in the primordial cosmology and in later time cosmology is affected by inhomogeneous perturbation calculations is not yet fully understood. This means that we have to be cautious about applying computer codes such as CMBFAST and CAMB, which were designed exclusively to give fits to data based on the standard $\Lambda$CDM model, to the more general inhomogeneous cosmology models.

The problem of the age of the universe and the amplification of perturbation calculations of structure growth and correlation functions have been addressed here. We found that as already discovered 1n one of the first phenomenological investigations of the inhomogeneous void cosmology~\cite{Moffat,Moffat2}, based on the exact pressureless LBT model, the age of the universe problem can be solved in these models and that an increased amplification and early onset of structure growth is to be expected.

An alternative inhomogeneous cosmology model based on the Szekeres-Szafron~\cite{Szafron} exact solution of Einstein's field equations has been investigated~\cite{Moffat9}. The solution does not in general have any symmetry and incorporates pressure as well as energy density. A particular cylindrically symmetric Szafron solution was studied, which can describe an early universe with non-uniform inflation, has homogeneous LFRW behavior at the surface of last scattering and can become inhomogeneous at a late time of the expansion of the universe.

{\bf Acknowledgements:}  I thank Bruce Bassett and Viktor Toth for stimulating discussions and Viktor Toth for a numerical calculation. This research was supported by a grant from NSERC. The Perimeter Institute is supported in part by the Government of Canada through NSERC and by the Province of Ontario through MEDT.

\end{document}